\documentclass[a4paper]{article}

\usepackage{INTERSPEECH2021}
\usepackage{cite}
\usepackage{float}
\usepackage{multirow}
\usepackage{url}
\usepackage{CJKutf8}
\usepackage{graphicx}
\usepackage{epstopdf}

\title{EchoFilter: End-to-End Neural Network for Acoustic Echo Cancellation}
\name{Lu Ma, Song Yang, Yaguang Gong, Xintian Wang, Zhongqin Wu\thanks{This work was supported by National Key R$\&$D Program of China, under Grant No. 2020AAA0104500. The corresponding author is Lu Ma. Email: malu6@tal.com, iamroad@163.com}}
\address{TAL Education Group, Beijing, China}
\email{\{malu6,yangsong1,gongyaguang,wangxintian,wuzhongqin\}@tal.com}

\begin{document}

\maketitle
\begin{abstract}
Acoustic Echo Cancellation (AEC) whose aim is to suppress the echo originated from acoustic coupling between loudspeakers and microphones, plays a key role in voice interaction. Linear adaptive filter (AF) is always used for handling this problem. However, since there would be some severe effects in real scenarios, such nonlinear distortions, background noises, and microphone clipping, it would lead to considerable residual echo, giving poor performance in practice. In this paper, we propose an end-to-end network structure for echo cancellation, which is directly done on time-domain audio waveform. It is transformed to deep representation by temporal convolution, and modelled by Long Short-Term Memory (LSTM) for considering temporal property. Since time delay and severe reverberation may exist at the near-end with respect to the far-end, a local attention is employed for alignment. The network is trained using multitask learning by employing an auxiliary classification network for double-talk detection. Experiments show the superiority of our proposed method in terms of the echo return loss enhancement (ERLE) for single-talk periods and the perceptual evaluation of speech quality (PESQ) score for double-talk periods in background noise and nonlinear distortion scenarios.
\end{abstract}

\section{Introduction}
\label{sec:intro}

Acoustic echo will arise when the microphone at the near-end coupling the loudspeaker's signal and its reverberation. This echo could be listened by the speaker itself at the far-end, degrading user feelings in speech and voice applications. Great attention has been payed to for decades \cite{Sondhi,Benesty}. Since there is a reference signal named the far-end, adaptive filters are always employed for estimating acoustic echo, thus achieving acoustic echo cancellation (AEC) by subtracting the estimation from the mixture \cite{Breining}. Several classical algorithms based on adaptive filtering (AF) have been proposed, such as least mean square (LMS) \cite{FLMS}, normalized LMS (NLMS) \cite{NLMS}, block LMS (BLMS) \cite{BLMS}, and etc. Among them the NLMS algorithm family is most widely used due to its relatively robust performance and low complexity, such as the frequency domain block adaptive filter (FDBAF) \cite{PBFDAF} used in WebRTC and the multidelay block frequency domain (MDF) \cite{MDF} adaptive filter used in Speex.

However, nonlinear distortion may be introduced to the acoustic echo in practices which are mostly caused by non-linear components on the audio devices, resulting in considerable residual echo. To overcome this problem, numeric methods have been proposed, such as Nonlinear AEC (NLAEC) method by using a set of nonlinear basis functions for echo estimation \cite{Stenger,Kuech,Carini} and nonlinear post-filtering method by cascading an additional nonlinear processing module for residual echo suppression \cite{Gustafsson,Turbin,Bendersky,Schwarz}. Recently, since its great potential in speech processing tasks, neural network (NN) has been used for echo cancellation or suppression, such as NN-based post-filtering method where NN is used for residual echo suppression instead of conventional post-filters \cite{MaLu,Halimeh}, NN-based NLAEC method where NN is used for modeling nonlinear echo \cite{Halimeh1}, separation-based method where source separation with an additional information from the far-end \cite{DeLiang} is used for AEC, NN-based adaptive filtering method where the structure of conventional linear filter is adopted for designing network \cite{Fazel,Fazel1}.

In this paper, we propose an end-to-end framework for acoustic echo cancellation. It is directly done on time-domain audio waveform. 1--D convolution and its transposed one are used for waveform encoding and decoding. Then a mask network constructed by Long Short-Term Memory (LSTM) layers for considering temporal property are designed for echo cancellation. Since time delay and severe reverberation may exist at the near-end with respect to the far-end, a local attention~\cite{attention,attention1} is employed for alignment. As an auxiliary task, a classification network designed for double-talk detection is embedded. It could give some useful information, such as only silence exists, only the near-end exits, only the far-end exists, or double-talk exists, prompting the mask network to make better echo cancellation. This idea is different from that of \cite{Fazel} where an auxiliary task of estimating echo is designed to improve the main task of estimating the near-end speech.
Additionally, our method is also inspired by \cite{VoiceFilter} where an embedding of the targeted speaker from a clean enrolled sequence is used for isolating that speaker in a mixture. While, no speaker encoder is needed here.

The remainder of this paper is organized as follows. Section 2 provides the details of our proposed network structure. Section 3 presents our experimental results. And finally, the summarization and discussions are given in Section 4.

\section{Network Architecture}
\label{sec:structure}

\begin{figure}[htb]
\centering
\includegraphics[scale=0.9]{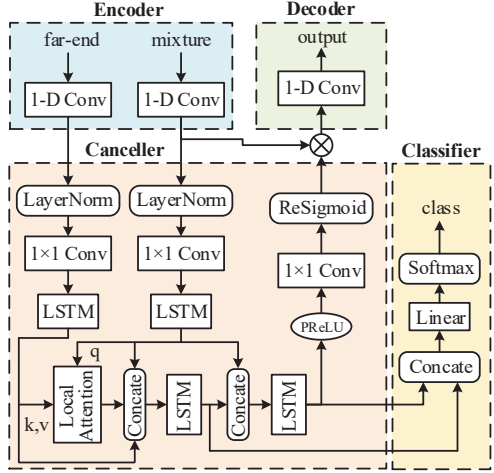}
\caption{Network architecture.}
\label{fig:structure}
\end{figure}

The network of acoustic echo cancellation is shown in Fig.~\ref{fig:structure}, which is constituted by four modules, i.e., encoder, canceller, classifier and decoder. Encoder modules are used to transform short segments of the mixture waveform and the far-end waveform into their corresponding representations in an intermediate feature space. These representations are then used to estimate a multiplicative function (mask) for the near-end signal at each time step. As an auxiliary task, the intermediate representations of the canceller module are used for double-talk detection. The near-end waveform is then reconstructed by transforming the masked representation using a decoder module.

\begin{figure}[htb]
\vspace{-0.2cm}
\centering
\includegraphics[scale=1.0]{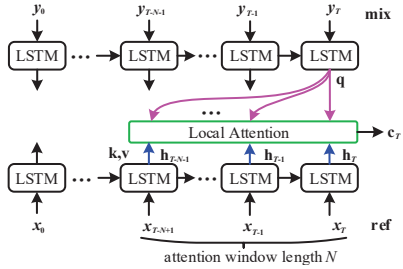}
\caption{Local attention for alignment.}
\label{fig:localattn}
\vspace{-0.4cm}
\end{figure}

\subsection{Encoder and Decoder}
\label{sec:encoder}

The encoder performs like short-time Fourier transform (STFT) with high-resolution frequency decomposition. The input audio is divided into overlapping segments of length $L$, represented by $x_{k} \in \mathbb{R}^{1 \times L}$, where $k=1, \ldots, \widehat{T}$ denotes the segment index and $\widehat{T}$ denotes the total number of segments. $x_{k}$ is transformed into a $N$-dimensional representation, by a 1--D convolution operation $\mathbf{w} \in \mathbb{R}^{1 \times N}$ (denoted by $1-D$ $Conv$), formulated by a matrix multiplication as ${\mathbf{w}}=\mathcal{H}(\mathrm{\mathbf{x}} \mathbf{U})$, where $\mathbf{U} \in \mathbb{R}^{N \times L}$ contains $N$ vectors (encoder basis functions) with length $L$ for each, $\mathcal{H}(\cdot)$ is the rectified linear unit (ReLU) function~\cite{ref_relu} to ensure non-negative of the representation.

The decoder reconstructs the waveform from this representation using a 1--D transposed convolution operation, which can be reformulated as another matrix multiplication as $\hat{\mathbf{x}}=\mathbf{w} \mathbf{V}$, where $\hat{\mathbf{x}} \in \mathbb{R}^{1 \times L}$ is the reconstruction of $\mathbf{x}$ and the rows in $\mathbf{V} \in \mathbb{R}^{N \times L}$ are the decoder basis functions, each with length $L$. The overlapping reconstructed segments are summed together
to generate the final waveforms.

\subsection{Canceller}
\label{sec:masking}

It is a mask estimation network. Layer normalization is firstly used to ensure that the calculation is invariant to the input scaling, which is a cumulative layer normalization (cLN) designed for causal configuration~\cite{ref_convtasnet}. The pointwise convolution ($1$$\times $$1$-$Conv(\cdot)$)~\cite{ref_depthwise} is used here as a bottleneck to compress the input channels, obtaining higher dimensional representation.

For each path of the mixture and the far-end, a LSTM layer is used after the pointwise convolution for temporal property modelling. Then, a local attention~\cite{attention,attention1} is used for time alignment, by computing the similarities between the mixture and the far-end with a look backward window as is shown in Fig.~\ref{fig:localattn}. It computes similarities between the mixture at the $T$-th frame and the far-end at the $T$-th frame together with the previous $N$$-1$ frames, obtaining a aligned far-end at the $T$-th frame. The mixture representation acts as the query (q), and the far-end representation acts as the key (k) and the value (v).
As is used in \cite{VoiceFilter} where LSTM is used to removing the interfering speaker, a LSTM layer is employed here for echo estimation by concatenating the mixture, the far-end and the aligned far-end. Then, another LSTM layer is used for separating the estimated echo from the mixture by concatenating the estimated echo and the mixture, obtaining a estimated near-end. This manipulation is inspired by the conventional adaptive filtering method where time alignment is firstly done followed by a linear filter for echo estimation, and then subtracted from the mixture.

Finally, the estimated near-end representation is mapped to masks by $1$$\times $$1$-$Conv(\cdot)$ with rectified sigmoid (ReSigmoid) activation function. It is defined by the multiplication of the ReLU and the Sigmoid, formulated by
\begin{equation} \label{eq:resigmoid}
\setlength{\abovedisplayskip}{3pt}
\setlength{\belowdisplayskip}{3pt}
x = relu(x)\times sigmoid(x)
\end{equation}

The $1$$\times $$1$-$Conv(\cdot)$ converts the number of channels back to that of the encoder. Before masking, the parametric rectified linear unit (PReLU)~\cite{prelu} is used to activate the representation.

\subsection{Classifier}
\label{sec:classification}

It is constituted by linear layer with softmax activation function, whose input is the concatenation of the estimated echo and the estimated near-end. It performs like double-talk detection, and could give some useful information, such as only silence exists, only the near-end exits, only the far-end exists, or double-talk exists, prompting the network to make better estimation. Vice versa, as long as the near-end representation is estimated precisely, could the double-talk be detected accurately.

\subsection{Loss Function}
\label{sec:loss_cun}
The loss function is obtained by combining the mean square error (MSE) of the estimated near-end waveform and the cross-entropy of the classification, formulated by,

\begin{equation} \label{eq:mask_loss}
\setlength{\abovedisplayskip}{-4pt}
\setlength{\belowdisplayskip}{7pt}
\left\{\begin{array}{l}
L_{total}=(1-\alpha) \cdot {Loss_{\rm{MSE}}} +\alpha \cdot {Loss_{\rm{CE}}} \\
{Loss_{\rm{MSE}}} = \sum_{n}\left(\hat{s}(n)-s(n)\right)^{2} \\
{Loss_{\rm{CE}}} = -\sum_{n}\sum_{i=1}^{i=M}p(\hat{c_i}(n))log(p({c_i}(n)))
\end{array}\right.
\end{equation}
where ${Loss_{\rm{MSE}}}$ is the MSE between the estimated near-end waveform $\hat{s}(n)$ and the clean one ${s}(n)$, ${Loss_{\rm{CE}}}$ is the cross-entropy between the estimated classification $\hat{c_i}(n)$ and the labelled classification ${c_i}(n)$, $i \in [1,2,..,M]$ is the the categories, $n$ is the time index, $\alpha \in[0,1]$ is the weighting factor.

\section{Experiments}
\label{sec:experiment}

\subsection{Performance metrics}
Two metrics are evaluated here, i.e., the echo return loss enhancement (ERLE) for single-talk periods (periods without near-end signal) and the perceptual evaluation of speech quality (PESQ) for double-talk periods. ERLE reveals the echo attenuation achieved by the system \cite{erle}, which is defined as
\begin{equation} \label{eq:resigmoid}
\setlength{\abovedisplayskip}{3pt}
\setlength{\belowdisplayskip}{3pt}
\mathrm{ERLE}=10 \log _{10}\left\{\frac{\mathcal{E}\left[y^{2}(n)\right]}{\mathcal{E}\left[\hat{s}^{2}(n)\right]}\right\}
\end{equation}
where $\mathcal{E}$ denotes the statistical expectation, $y(n)$ and $\hat{s}(n)$ are the mixture and estimated near-end at time index $n$ respectively.

PESQ$\footnote{\url{https://www.itu.int/rec/T-REC-P.862/en}}$ is obtained by comparing the estimated near-end $\hat{s}(n)$ with the raw one $s(n)$~\cite{pesq}.

\subsection{Data preparation}
\begin{figure}[htb]
\vspace{-0.2cm}
\centering
\includegraphics[scale=0.9]{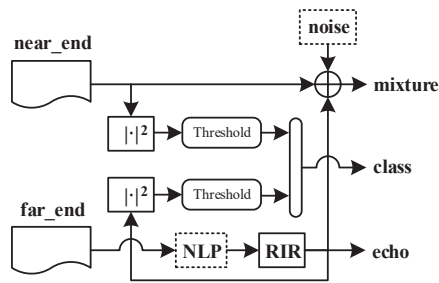}
\caption{Schematic of data preparation.}
\label{fig:data}
\vspace{-0.2cm}
\end{figure}

Since TIMIT dataset was widely used in literatures to evaluate AEC performance, we follows the data preparation method as is referred to in \cite{DeLiang,Halimeh}, resulting in 3500 training mixtures, and 300 test mixtures. The generation for inputs and labels are illustrated in Fig.~\ref{fig:data}. Here, the RIR (Room Impulse Response) is used for simulating room reflections~\cite{rir}, and the NLP (Non-Linear Processing) is used for emulating the distortion introduced by loudspeaker. The far-end waveform is first distorted by the NLP, and then convoluted with the RIR, obtaining the echo waveform received at microphone. It is then added to the near-end waveform to obtain the mixture waveform received by microphone. The mixture is then added by a noise waveform.

Meanwhile, the signals' energy of the near-end and the echo are calculated and compared to the threshold, judging whether there is voice at the corresponding branch. Four categories are obtained, i.e., no voice exist at both branches (i.e., silence), only voice exists at the near-end (i.e., near-end single-talk), only voice exists at the far-end (i.e., far-end single-talk), voice exist at both ends (i.e., double-talk), acting as classification labels.

The RIR in the experiments are configured as is referred to in \cite{DeLiang,Halimeh}, resulting in 7 RIRs, of which the first 6 RIRs are used to generate training mixtures and the last one is used to generate test mixtures. The hard clipping is used to simulate the power amplifier of loudspeaker, and the memoryless sigmoidal function is applied to emulate the nonlinear characteristic of loudspeaker, resulting in $x_{nl}(n)$ for nonlinear inputs.

Finally, the linear model $y_{lin}(n)$ and nonlinear model $y_{nl}(n)$ of acoustic path are obtained by convolving the linear input $x(n)$ and the non-linear input $x_{nl}(n)$ with a randomly chosen RIR $g(n)$ as,
\begin{equation}
\setlength{\abovedisplayskip}{3pt}
\setlength{\belowdisplayskip}{3pt}
\begin{array}{l}
y_{lin}(n) = x(n) * g(n),\quad y_{nl}(n) = x_{nl}(n) * g(n)
\end{array}
\end{equation}
where $*$ indicates the convolution manipulation.

For training mixtures, we generated the mixtures at signal to echo ratio (SER)\cite{DeLiang} level randomly chosen from {-6, -3, 0, 3, 6} dB. For testing mixtures, we generated the mixtures at three different SER levels {0, 3.5, and 7} dB.

\subsection{Model configurations}
In our experiments, waveforms at 16 kHz sample rate are directly served as the inputs. The network structure is depicted in Fig.~\ref{fig:structure} with the configuration listed in Table~\ref{tab:configuration}. The symbols in the table are defined in Section \ref{sec:encoder}. The decoder was chosen as the transposed version of the encoder, and is denoted by $1$--$D$ $ConvT$ in the table. All the hidden size of the LSTM layer are same. All the output channel of the $1$--$D$ $Conv$ are same, expect for the last one which is used for converting the channel back to that of the encoder's output. $\emph{Adm}$ algorithm was used for training with an exponential learning rate decaying strategy, where the learning rate starts at $1 \times 10^{-4}$ and ends at $1 \times 10^{-8}$. The total number of epochs was set to be 200. The weighting factor in Eq. (\ref{eq:mask_loss}) was $\alpha = 0.001$. The criteria for early stopping is no decrease in the loss function on validation set for 10 epochs. We found that it was appropriate to chose 100 as window size of the local attention in the experiments for accommodating to large reverberation time.

\begin{table}[htb]
\setlength{\abovecaptionskip}{0.1cm}
\setlength{\belowcaptionskip}{-0.1cm}
\centering
\caption{Network configurations}
\label{tab:configuration}
\begin{tabular}{|l|l|l|l|l|l|l|}
\hline
\textbf{Module} & \textbf{Structure} & \textbf{Configuration} \\ \hline \hline
{Encoder} & $1$-$D$ $Conv$ & $N=512$, $L=160$\\ \hline
{Decoder} & $1$-$D$ $ConvT$ & $N=512$, $L=160$\\ \hline
\multirow{3}{*}{Canceller} & $1$$\times $$1$-$Conv$ & output size: 256 \\ \cline{2-3}
 & LSTM & hidden size: 256 \\ \cline{2-3}
 & Attention & window size: 100 \\ \hline
 {Classifer} & Linear & hidden size: 4 \\ \hline
\end{tabular}
\vspace{-10pt}
\end{table}

\subsection{Results}

We first evaluated our method using linear model of acoustic path. Three schemes were used for comparisons, i.e., frequency domain normalized least mean square (denoted by AES+RES) \cite{fdnlms},  bidirectional long short-term memory method (denoted by BLSTM) \cite{DeLiang}, deep multitask (denoted by Multitask) \cite{Fazel}. Table~\ref{tab:linear} shows the average ERLE values and PESQ gains for these schemes. The $\Delta$PESQ is calculated as the difference of PESQ value of each method with respect to its unprocessed PESQ. The results show that our end-to-end method obtain both higher ERLE and PESQ performance compared with Multitask method. It also shows that the proposed method outperforms both AES+RES and BLSTM methods in all conditions. Audio samples randomly selected from the experiments can be found in our GitHub repository$\footnote{\url{https://github.com/ROAD2018/end_to_end_aec}}$.

\begin{table}[htb]
\setlength{\abovecaptionskip}{0.1cm}
\setlength{\belowcaptionskip}{-0.1cm}
\centering
\caption{Performance of simulated linear model}
\label{tab:linear}
\begin{tabular}{|c|c|c|c|c|}
\hline
\multirow{2}{*}{\textbf{Metrics}}     & \multirow{2}{*}{\textbf{Methods}} & \multicolumn{3}{c|}{\textbf{Testing SER (dB)}} \\ \cline{3-5}
                      &       & \textbf{0}     & \textbf{3.5}    & \textbf{7}           \\ \hline \hline
\multirow{4}{*}{\textbf{ERLE (dB)}} & AES+RES                 & 29.38       & 25.88       & 21.97       \\ \cline{2-5}
                      & BLSTM    & 51.61   & 50.04  & 47.42       \\ \cline{2-5}
                      & Multitask   & 64.66    & 64.16   & 62.26    \\ \cline{2-5}
                      & Proposed   & 78.28   & 76.06   &  71.79     \\ \hline
\multirow{4}{*}{\textbf{$\Delta$ PESQ}} & AES+RES & 0.93 & 0.81 & 0.68 \\ \cline{2-5}
                      & BLSTM      & 0.8    & 0.78   & 0.74        \\ \cline{2-5}
                      & Multitask   & 1.04  & 1.02   & 0.99      \\ \cline{2-5}
                      & Proposed    & 1.63  & 1.66   & 1.59      \\ \hline
\end{tabular}
\vspace{-2pt}
\end{table}

Further, nonlinear model of acoustic path for our proposed method is also evaluated. In this experiment, we used $y_{nl}$ for generating the echo signals, bringing both power amplifier clipping and loudspeaker distortions to the echo signals. Two schemes were compared, i.e., AES+RES method and the Multitask method. As is shown in Table~\ref{tab:nonlinear}, our end-to-end scheme gain the best performance both for ERLE and PESQ.

\begin{table}[htb]
\setlength{\abovecaptionskip}{0.1cm}
\setlength{\belowcaptionskip}{-0.1cm}
\centering
\caption{Performance of simulated nonlinear model}
\label{tab:nonlinear}
\begin{tabular}{|c|c|c|c|c|}
\hline
\multirow{2}{*}{\textbf{Metrics}}   & \multirow{2}{*}{\textbf{Methods}} & \multicolumn{3}{c|}{\textbf{Testing SER (dB)}} \\ \cline{3-5}
                      &       & \textbf{0}     & \textbf{3.5}    & \textbf{7}           \\ \hline \hline
\multirow{3}{*}{\textbf{ERLE (dB)}} & AES+RES & 16.76 & 14.26 & 12.33 \\ \cline{2-5}
                      & Multitask     & 61.79   & 60.52   & 59.47  \\ \cline{2-5}
                      & Proposed     &  77.29   & 74.32  & 71.37  \\ \hline
\multirow{3}{*}{\textbf{$\Delta$ PESQ}} & AES+RES & 0.54 & 0.43 & 0.31 \\ \cline{2-5}
                      & Multitask       & 0.84   & 0.83    & 0.81  \\ \cline{2-5}
                      & Proposed       &  1.53   & 1.59   & 1.54      \\ \hline
\end{tabular}
\end{table}

Additionally, we also evaluated the performance in presence of additive noise and nonlinear model of acoustic path. When generating the training data, we added a white noise at 10dB SNR level with nonlinear acoustic path at 3.5dB SER level. We compared our method against a conventional AES+RES method and the Multitask method. Our end-to-end framework gives the best performance as is shown in Table~\ref{tab:noisy}.

\begin{table}[htb]
\setlength{\abovecaptionskip}{0.1cm}
\setlength{\belowcaptionskip}{-0.1cm}
\centering
\caption{Performance for noisy scenario}
\label{tab:noisy}
\begin{tabular}{|c|c|c|c|c|}
\hline
{\textbf{Metrics}} & {\textbf{Methods}}
& {\textbf{Value}} \\ \hline \hline
\multirow{3}{*}{\textbf{ERLE (dB)}} & AES+RES    & 10.13      \\ \cline{2-3}
                      & Multitask     & 46.12    \\ \cline{2-3}
                      & Proposed     & 59.32      \\ \hline
\multirow{3}{*}{\textbf{$\Delta$ PESQ}} & AES+RES   & 0.21      \\ \cline{2-3}
                      & Multitask       & 0.70    \\ \cline{2-3}
                      & Proposed       & 1.30    \\ \hline
\end{tabular}
\vspace{-4pt}
\end{table}

\begin{figure}[htb]
	\begin{minipage}[b]{1.0\linewidth}
		\centering
		\centerline{\includegraphics[width=7.5cm, height=0.9cm]{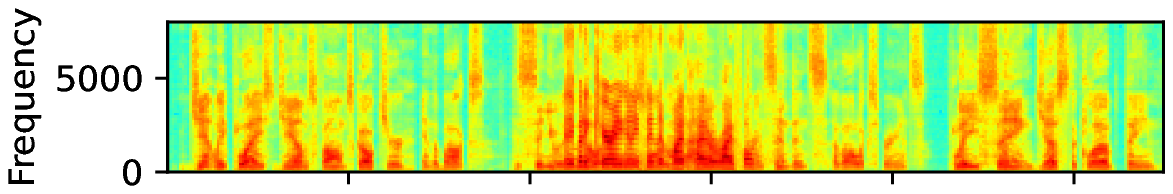}}
        \vspace{-0.05cm}
	\end{minipage}
	\begin{minipage}[b]{1.0\linewidth}
		\centering
		\centerline{\includegraphics[width=7.5cm, height=0.9cm]{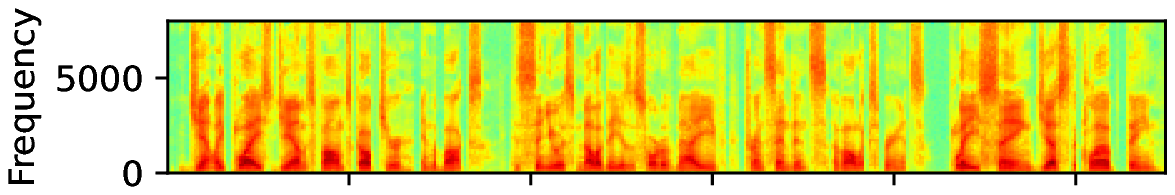}}
        \vspace{-0.05cm}
	\end{minipage}
	\begin{minipage}[b]{1.0\linewidth}
		\centering
		\centerline{\includegraphics[width=7.5cm, height=0.9cm]{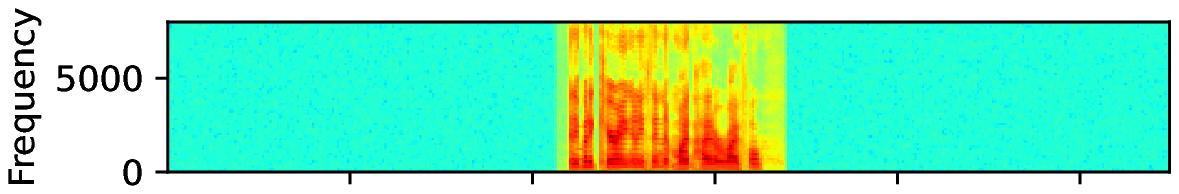}}
        \vspace{-0.05cm}
	\end{minipage}
	\begin{minipage}[b]{1.0\linewidth}
		\centering
		\centerline{\includegraphics[width=7.5cm, height=0.9cm]{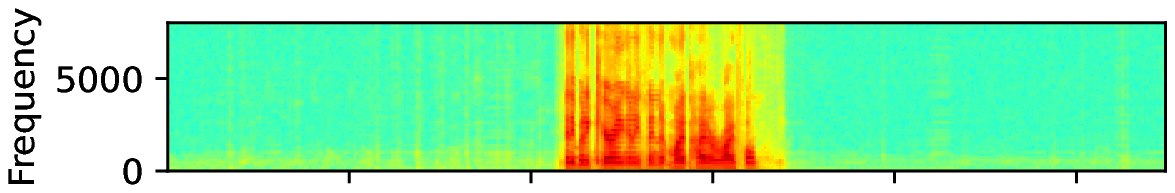}}
        \vspace{-0.05cm}
	\end{minipage}
	\begin{minipage}[b]{1.0\linewidth}
		\centering
		\centerline{\includegraphics[width=7.5cm, height=0.9cm]{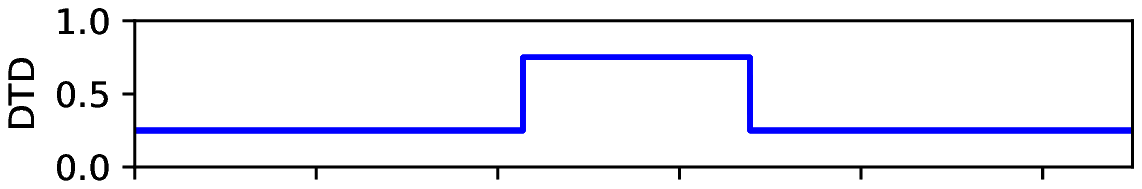}}
        \vspace{-0.05cm}
	\end{minipage}
	\caption{Spectrograms of microphone, far-end, near-end, estimated near-end and double-talk detection in nonlinear model of acoustic path and 0dB SER.}
	\label{fig:spectrogram}
\end{figure}

Spectrograms in Fig.~\ref{fig:spectrogram} illustrates an AEC example using our proposed method. In the figure, from the top to the bottom, there are the mixture, the far-end, the echo, the near-end, the estimated near-end and the double-talk detection. The x-axis and the y-axis are time and frequency respectively. It is obvious that, the double-talk detection matches the voice appearance of the near-end and the far-end perfectly. This may be a key point for our model to gain a larger ERLE value. Moreover, from experiments we found that the proposed ReSigmoid activation function is superior to the Sigmoid activation function for obtaining a higher echo suppression.
Meanwhile, the observation that good near-end waveform can be recovered may owe to the end-to-end architecture that is directly done on audio waveform.

From the aforementioned comparisons, we can conclude that our framework has better performance for synthesized data in terms of linear and non-linear acoustic models. Moreover, to evaluate it for real scenarios, real measured RIRs selected from the Aachen impulse response database \cite{realRIR} were used in the experiments for evaluating real-world performance. In this experiment, we used the real measured RIRs captured in ``office'', ``meeting room'', ``lecture room'', ``stairway1'', ``stairway2'', ``bathroom'', and ``lecture room'' for training and ``corridor'' for testing in HHP. Two schemes were compared as is listed in Table~\ref{tab:reallinear}, i.e., the AES+RES method and the context-aware method (denoted by CAD-AEC) \cite{Fazel1}. Our method gains higher ERLE and PESQ values compared with the AES+RES and the CAD-AEC methods. Especially for the ERLE gain, this owes to the auxiliary network for double-talk detection.

\begin{table}[htb]
\setlength{\abovecaptionskip}{0.1cm}
\setlength{\belowcaptionskip}{-0.1cm}
\centering
\caption{Performance of real measured linear model}
\label{tab:reallinear}
\begin{tabular}{|c|c|c|c|c|}
\hline
\multirow{2}{*}{\textbf{Metrics}}   & \multirow{2}{*}{\textbf{Methods}} & \multicolumn{3}{c|}{\textbf{Testing SER (dB)}} \\ \cline{3-5}
                      &       & \textbf{0}     & \textbf{3.5}    & \textbf{7}           \\ \hline \hline
\multirow{3}{*}{\textbf{ERLE (dB)}} & AES+RES    & 12.16    & 11.46    & 10.52       \\ \cline{2-5}
                      & CAD-AEC     & 56.51   & 60.49   & 61.39  \\ \cline{2-5}
                      & Proposed    & 70.93   & 69.96   & 66.89        \\ \hline
\multirow{3}{*}{\textbf{$\Delta$ PESQ}} & AES+RES   & 0.57        & 0.53    & 0.48       \\ \cline{2-5}
                      & CAD-AEC       & 1.11   & 1.06    & 1.00  \\ \cline{2-5}
                      & Proposed      & 1.23   & 1.35    & 1.31      \\ \hline
\end{tabular}
\vspace{-5pt}
\end{table}

Finally, to evaluate the performance when the training and testing conditions are more different than the previous experiments, we generated seven synthetic RIRs for training and again tested on data that was created by the real measured ``corridor'' RIRs. The reverberation time ($T_{60}$) is matched with the `corridor'' environment which is selected from $\{0.2, 0.4, 0.6, 0.8, 0.9, 1.0, 1.25\}$s. The deployment for the `corridor'' environment is referred to the Fig. 9(c) in \cite{realRIR}. For nonlinear RIRs, we used $y_{nl}$ for generating the echo signals, therefore our nonlinear model contains both power amplifier clipping and loudspeaker distortions. We again compared the results of our method against the CAD-AEC method \cite{Fazel1} both in linear and nonlinear RIRs.
As is revealed in~Table \ref{tab:generalization}, for both linear and nonlinear RIR models, our end-to-end method outperforms the CAD-AEC method in terms of the ERLR and the PESQ. This superiority is more obvious for the ERLE gain. It also can be found that the PESQ gain with respect to the CAD-AEC method increases with the increasing of SER.

\begin{table}[htb]
\setlength{\abovecaptionskip}{0.1cm}
\setlength{\belowcaptionskip}{-0.1cm}
\centering
\caption{Performance of training on synthetic RIRs and testing on real RIRs.}
\label{tab:generalization}
\begin{tabular}{|c|c|c|c|c|}
\hline
\multirow{2}{*}{\textbf{Metrics}}   & \multirow{2}{*}{\textbf{Methods}} & \multicolumn{3}{l|}{\textbf{Testing SER (dB)}} \\ \cline{3-5}
                      &       & \textbf{0}     & \textbf{3.5}    & \textbf{7}           \\ \hline \hline
\multicolumn{5}{|c|}{\textbf{Linear RIR Model}}  \\ \hline \hline
\multirow{2}{*}{\textbf{ERLE}} & CAD-AEC & 42.66   & 47.96   & 52.47  \\ \cline{2-5}
                      & Proposed  & 68.71     & 66.49     & 65.11        \\ \hline
\multirow{2}{*}{\textbf{PESQ}} & CAD-AEC   & 2.76   & 2.92    & 3.06  \\ \cline{2-5}
                      & Proposed  & 2.95   & 3.27   & 3.58    \\ \hline \hline
\multicolumn{5}{|c|}{\textbf{Nonlinear RIR Model}}  \\ \hline \hline
\multirow{2}{*}{\textbf{ERLE}} & CAD-AEC & 19.08   & 19.97   & 21.64  \\ \cline{2-5}
                      & Proposed  & 66.39  & 64.97  & 65.27    \\ \hline
\multirow{2}{*}{\textbf{PESQ}} & CAD-AEC  & 2.74   & 2.93    & 3.09  \\ \cline{2-5}
                      & Proposed  & 2.76   & 3.13   & 3.45     \\ \hline
\end{tabular}
\vspace{-4pt}
\end{table}

\vspace{-4pt}
\section{Conclusions}

An end-to-end framework is proposed for acoustic echo cancellation. It is directly done on audio waveform. 1--D Temporal convolution is used for transforming the waveform into deep representation, followed by LSTM layers which is used for modeling the temporal property. Meanwhile, local attention is employed to make alignment between the mixture and the far-end. The network is trained using multitask learning by employing an auxiliary classification network for double-talk detection, giving some useful information and prompting the network to make better estimation.  Experimental results show the effectiveness and superiority of the proposed framework in double-talk, background noise, and nonlinear distortion scenarios. Moreover, it can be implemented in real-time scenarios.

\bibliographystyle{IEEEtran}

\bibliography{mybib}

\begin{thebibliography}{9}
 \bibitem[1]{Sondhi}
   M.\ Sondhi,
   ``An adaptive echo canceller,''
   \textit{Bell Syst. Tech. J.}, vol.~46, no.~3, pp.~497--511, 1967.

 \bibitem[2]{Benesty}
   J.\ Benesty and T.\ G$\ddot{a}$nsler,
   ``Advances in network and acoustic echo cancellation,''
   \textit{Advances in network and acoustic echo cancellation}, Springer, 2001.

 \bibitem[3]{Breining}
   C.\ Breining and et al., ``Acoustic echo control -- an application of very-highorder adaptive filters,'' \textit{IEEE Signal Process. Magazine}, vol.~16, no.~4, pp.~42--69, 1999.

 \bibitem[4]{FLMS}
   E.\ Ferrara, ``Fast implementations of LMS adaptive filters,'' \textit{IEEE Transactions on Acoustics, Speech, and Signal Processing,}, vol.~28, no.~4, pp.~474--475, 1980.

 \bibitem[5]{NLMS}
   R.\ Tyagi R.\ Singh and R.\ Tiwari, ``The performance study of NLMS algorithm for acoustic echo cancellation,'' \textit{International Conference on Information, Communication, Instrumentation and Control,}, 2017, pp.~1--5, Indore.

 \bibitem[6]{BLMS}
   G.\ A.\ Clark, S.\ K.\ Mitra, and S.\ R.\ Parker, ``Block implementation of adaptive digital filters,'' \textit{IEEE Transactions on Acoustics, Speech, and Signal Processing}, vol. ASSP-29, pp.~744--752, 1981.

 \bibitem[7]{PBFDAF}
   P. Borrallo Jos$\acute{e}$M. and M.\ G.\ Otero, ``On the implementation of a partitioned block frequency domain adaptive filter (PBFDAF) for long acoustic echo cancellation,''
   \textit{Signal Processing,}, vol.~27, no.~3, pp.~301--315, 1992.

 \bibitem[8]{MDF}
   J.\ S.\ Soo, K.\ K.\ Pang, ``Multidelay block frequency domain adaptive filter,'' \textit{IEEE Transactions on Acoustics, Speech, and Signal Processing}, vol.~38, no.~2, pp.~373--376, 1990.

 \bibitem[9]{Stenger}
   A.\ Stenger and W.\ Kellermann, ``Adaptation of a memoryless preprocessor for nonlinear acoustic echo cancelling,'' \textit{Signal Process}, vol.~80, pp.~1747--1760, 2000.

 \bibitem[10]{Kuech}
   F.\ Kuech et al., ``Nonlinear acoustic echo cancellation using adaptive orthogonalized power filters,'' \textit{IEEE International Conference on Acoustics, Speech, and Signal Processing}, Philadelphia, PA, USA, 2005, pp.~105--108.

 \bibitem[11]{Carini}
   A.\ Carini et al., ``Introducing Legendre nonlinear filters,'' \textit{IEEE International Conference on Acoustics, Speech, and Signal Processing}, Florence, Italy, May 2014, pp.~7939--7943.

 \bibitem[12]{Gustafsson}
   S.\ Gustafsson, R.\ Martin, and P.\ Vary, ``Combined acoustic echo control and noise reduction for hands-free telephony,'' \textit{Signal Process}, vol.~64, no.~1, pp.~21--32, 1998.

 \bibitem[13]{Turbin}
   V.\ Turbin, A.\ Gilloire, and P.\ Scalart, ``Comparison of three postfiltering algorithms for residual acoustic echo reduction,'' \textit{IEEE International Conference on Acoustics, Speech, and Signal Processing}, 1997, pp.~307--310.

 \bibitem[14]{Bendersky}
   D.\ A.\ Bendersky, J.\ W.\ Stokes, and H.\ S.\ Malvar, ``Nonlinear residual acoustic echo suppression for high levels of harmonic distortion,'' \textit{IEEE International Conference on Acoustics, Speech, and Signal Processing}, 2008, pp.~261--264.

 \bibitem[15]{Schwarz}
   A.\ Schwarz, C.\ Hofmann, and W.\ Kellermann, ``Spectral feature based nonlinear residual echo suppression,'' \textit{in Applications of Signal Processing to Audio and Acoustics (WASPAA), 2013 IEEE Workshop on}, 2013, pp.~1--4.

 \bibitem[16]{MaLu}
   L.\ Ma, H.\ Huang, P.\ Zhao, and et al., ``Acoustic echo cancellation by combining adaptive digital filter and recurrent neural network,'' \textit{arXiv preprint}, arXiv:2005.09237, 2020.

 \bibitem[17]{Halimeh}
   M. M. Halimeh, T. Haubner, A. Briegleb, A. Schmidt, W. Kellermann, ``Combining adaptive filtering and complex-valued deep postfiltering for acoustic echo cancellation,'' \textit{ResearchGate preprint}, DOI: 10.13140/RG.2.2.14083.94241, avaiable:\textit{https://www.researchgate.net/publication/345003111}, 2020.

 \bibitem[18]{Halimeh1}
   M.\ M.\ Halimeh, C.\ Huemmer, W.\ Kellermann, ``A neural network-basedn nonlinear acoustic echo canceller,'' \textit{IEEE Signal Processing Letters}, 2019, vol. PP, no. 99, pp.~1--1.

 \bibitem[19]{DeLiang}
   H. Zhang, D. L. Wang, ``Deep learning for acoustic echo cancellation in noisy and double-Talk scenarios,'' \textit{Interspeech}, 2018, pp.~3239--3243.

 \bibitem[20]{Fazel}
   A. Fazel, M. El-Khamy, J. Lee, ``Deep multitask acoustic echo cancellation,'' \textit{Interspeech}, 2019, pp.~4250--4254.

 \bibitem[21]{Fazel1}
   A. Fazel, M. El-Khamy, J. Lee, ``CAD-AEC: context-aware deep acoustic echo cancellation,'' \textit{ICASSP}, 2020, pp.~6919--6923.

 \bibitem[22]{attention}
   R. W. Jack, R. Ali, ``Do transformers need deep long--range memory?'' \textit{Proceedings of the 58th Annual Meeting of the Association for Computational Linguistics}, 2020, pp.~7524--7529.

 \bibitem[23]{attention1}
   R. Aurko, S. Mohammad, V. Ashish, G. David, ``Efficient content-based sparse attention with routing transformers,'' \textit{arXiv preprint}, arXiv:2003.05997, 2020.

 \bibitem[24]{VoiceFilter}
   Q. Wang, H. Muckenhirn, K. Wilson and et al., ``VoiceFilter: targeted coice separation by speaker-conditioned spectrogram masking,'' \textit{Interspeech}, 2019.

 \bibitem[25]{ref_relu}
    K. Hara , D. Saito and H. Shouno , ``Analysis of function of rectified linear unit used in deep learning,'' \textit{International Joint Conference on Neural Networks (IJCNN)}, 2015, pp. 1--8.

  \bibitem[26]{ref_convtasnet}
    Y. Luo, N. Mesgarani , ``Conv-TasNet: surpassing ideal time--frequency magnitude masking for speech separation,'' \textit{IEEE/ACM Transactions on Audio, Speech, and Language Processing}, vol.~27, no.~8, pp.~1256--1266, 2019.

 \bibitem[27]{ref_depthwise}
    F. Chollet , ``Xception: deep learning with depthwise separable convolutions,'' \textit{IEEE Conference on Computer Vision and Pattern Recognition}, 2017, pp. 1800--1807.

 \bibitem[28]{prelu}
    K. He, X. Zhang, S. Ren, J. Sun, ``Delving deep into rectifiers: surpassing human-level performance on ImageNet classification,'' \textit{2015 IEEE International Conference on Computer Vision (ICCV)}, Santiago, Chile, 2015, pp. 1026--1034.


 \bibitem[29]{erle}
    G.\ Enzner, H.\ Buchner, A.\ Favrot, F. Kuech, ``Acoustic echo control,'' \textit{in Academic Press Library in Signal Processing,} Elsevier, 2014, vol.~4, pp.~807--877.

 \bibitem[30]{pesq}
    A.\ W.\ Rix, J.\ G.\ Beerends, M.\ P.\ Hollier, and et al., ``Perceptual evaluation of speech quality (pesq)-a new method for speech quality assessment of telephone networks and codecs,'' \textit{IEEE International Conference on Acoustics, Speech, and Signal Processing}, vol.~2. 2001, pp.~749--752.


 \bibitem[31]{rir}
    J.\ B.\ Allen, D.\ A.\ Berkley, ``Image method for efficiently simulating small--room acoustics,'' \textit{The Journal of Acoustic Society of America}, vol.~65, no.~4, pp.~943--950, 1979.

 \bibitem[32]{fdnlms}
    C.\ Faller, J.\ Chen, ``Suppressing acoustic echo in a spectral envelope space,'' \textit{IEEE Transactions on Acoustic, Speech and Signal Processing}, vol.~13, no.~5, pp.~1048--1062, 2005.

 \bibitem[33]{realRIR}
    M.\ Jeub, M.\ Sch$\ddot{a}$fer, H.\ Kr$\ddot{u}$ger, and et al., ``Do we need dereverberation for hand-held telephony?'' in \textit{Proceedings of 20th International Congress on Acoustics}, 2010, pp.~1--7.

 \end{thebibliography}


\end{document}